# Hybrid System for Solar Energy Conversion with Nano-Structured Electrodes

M.M. Nishchenko[1], M A Shevchenko[1], E A Tsapko[1], A A Frolov[2], G A Frolov[2], L. L. Sartinska[2], A. I. Blanovsky[3]

[1]Kurdumov Institute for Metal Physics, NAS of Ukraine,
[2]Frantsevich Institute for Problems of Materials Science, NAS of Ukraine
[3]Westside Environmental Technology (WETC), Los Angeles, CA, USA

**Abstract**

Solar energy is clean and renewable but has a low flux density around 1 kW/m$^2$. The lack of a continuous and reliable power supply reduces their average daily output to 3 – 7 kWh/m². Since the infrared part is not utilized to produce electricity, conventional photovoltaic (PV) cells use only about half of the solar spectrum. Direct radiation flux is also necessary for solar Stirling engine (SE) to work efficiently (33%). The hybrid system can largely overcome these problems.

The design combines the PV and energy storage-integrated SE or thermal field emission (TFE) technologies in the same concentrating solar power (CSP) system, providing a great potential in terms of energy production per unit area. The PV cells can be deposited on the CSP main mirror to allow the system to collect and convert the diffuse component of the light.

The design allows utilization of existing equipment, particularly a high-grade (around 700$^0$ C) parabolic dish or midgrade (around 300$^0$C) Fresnel concentrator for rooftop installation. To optimize the use of solar energy, beam splitters thermally decouple the modules by splitting the solar spectrum into three spectral ranges and directing visible, ultraviolet and infrared radiation into the PV cells, gate/Cs-filled gap and cavity-type solar receiver, respectively.

A photon-enhanced gate electrode creates on the cathode surface an electrostatic field large enough to compensate space charge field and initiates TFE process. In the TFE cathodes with nano-structured surfaces, the current density can reach values close to the field emission limit.

The design of the electrodes based on nano-structured emission materials were experimentally explored by the co-authors. In these experiments, efficiency of heat-to-electricity conversion was investigated and conditions for advanced nano-materials application for harvesting solar energy were found.

## 1. Introduction

Advances in nano-materials research are at the heart of solar power technologies. Particularly, nano-materials promise efficient and highly emissive thermionic surfaces – central to the design of highly efficient heat-to-electricity conversion devices. The proposed nano-material based TFE solar energy converter could perform at much lower temperatures than conventional thermionic (TE) converters. The converters have no highspeed moving parts, lower maintenance cost and be able to deliver reliable electricity production over a long service life.

Unlike the TE devices that operate at temperatures well above 1,500˚C, the TFE electrodes could operate at temperatures as low as 650 – 750˚C. This technology could be offered for industrial or domestic use and could be installed alongside or integrated into conventional solar power plants.

Recently, synthesis and characterization of nano-materials were performed to form electronically active CSP structures. Also, process of gas ionization under the influence of solar radiation at different conditions were analyzed.

## 2. Thermionic Experiment

It is well known that nano-tubes, nano wires, nano whiskers can serve as concentrators of the electric field and, therefore, they can be effectively used in the electrodes of thermo emission converters. This follows from the quantum nature of these effects in nano-structured materials. Moreover, the position of energy levels depends on interlayer spacing and structural defects.
Thin edges of nano-tubes with the radius of curvature r ~ 1-10 nm in the electric field of positive ions or gate electrodes will multiplied field for a few orders. Therefore, a potential barrier for electron (effect Schottky) and power losses will decrease significantly [1-6].
Samples of multi-walled carbon nanotubes (MWCNT) were synthesized by catalytic pyrolysis of ethylene. Catalysts $Al_3FeMo_{0,21}$ and $Al_3FeMo_{0,07}$ were used. A catalyst was mixed up with high dispersed pyrogenous silica (A-300) during 5 min in a mixer. Synthesis of multi-walled carbon nano-tubes was performed in a cylinder quartz reactor with the electro mechanic rotation of reactor. Obtained nano-tubes were pickled by mixture of the solution of ammonium bifluoride and muriatic acid for purification from mineral impurities.
As a result, two types of carbon nano-tubes have been produced: nano-tubes without defects and nano-tubes with dislocations. The mean diameter of the MWCNT is about 11 nm and interlayer spacing is about 3,5 nm. MWCNT can consist up to 10 graphene layers. Electron emission of samples composed of mixture of multi-walled carbon nano-tubes (MWCNT) and expanded graphite (EG) under concentrated solar radiation (in a mode of short circuit and in a mode of idling) has been studied. Recently, a solar energy conversion system was developed for investigation of nano-structured material emission properties at ~0,1 Pa vacuum. In the past conceptual experiments, solar radiation struck the 2 m diameter mirror, reflecting radiation through the quartz window of the vacuum chamber and heating the nanostructure cathode by focusing on the surface of the cathode as a spot with diameter of about 10mm (Fig. 1).

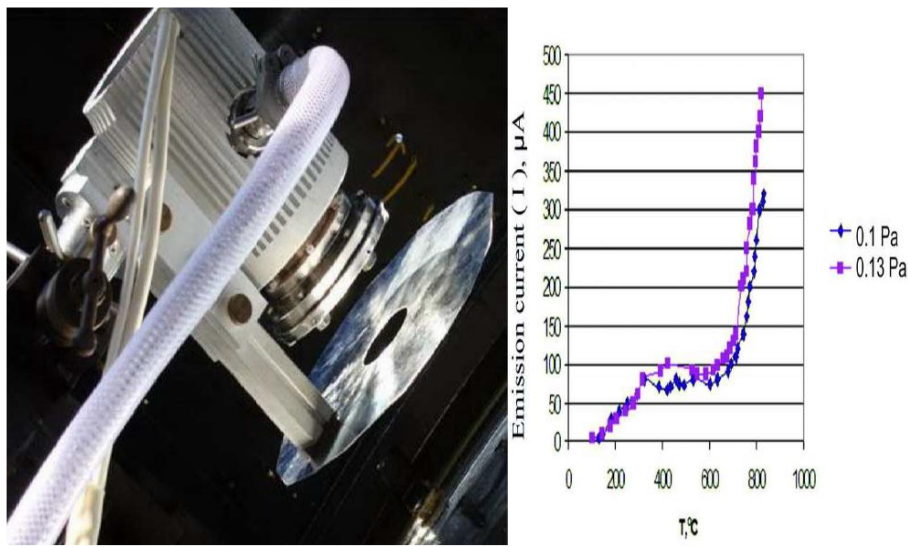

Fig. 1. (a) A vacuum chamber and (b) Emission current ( I ) versus a temperature of the cathode.

Nondestructive methods such as a positron annihilation spectroscopy was used to study the state and electron structure of nano-materials at a new level. Nano-structured material study is based on the tendency of positrons to get trapped in low electron density regions, where they annihilate with electrons emitting gamma rays, which carry unique information about atomic order defects in solids and free volumes through energy and time domain measurements. It was demonstrated by the method of the angular distribution of annihilation photons that defects of some MWCNT with the transverse size 0,5-0,6 nm, which equal to the doubled value of interlayer spacing can trap positrons. Such size of free volume can give the core of edge dislocation [7].

The ground and excited levels of dislocations are in subjacent positions in comparison with relevant levels of nearby (defect less) areas. It means that electrons are trapped by the core of edge dislocation. Therefore, according to data of electron-positron annihilation there is a redistribution of electrons and an increase the concentration of free electrons in the core of dislocation of nano-tubes. To improve system efficiency, computer modeling of the converter and simulation of various design configurations were performed.

The dependence of short circuit current from the temperature was measured. The highest value of current at the cathode surface of CNT-EG can attain 450 µA at the highest achieved temperature. It is possible to select three characteristic regions on the both curves of I=f (T) which show that process is many-staged. In the first stage there is an increase of current up to 100 µA in the range of temperatures from 300 to 350 ºC. This process is stipulated by the charge carrier's concentration increase as a result of increasing solar radiation intensity.

In the second stage there is a horizontal region on dependence I=f (T) which specifies the constant value of the anode's current (80-90 µA) for both curves in the range of temperatures from 350 to 650 ºC (Fig. 1b). There is an increase of current which achieves 450 µA at the temperature higher than 700 ºC obtained at the pressure of 0.13 Pa, that specifies the growth of charge carrier's concentration in an inter electrode space.

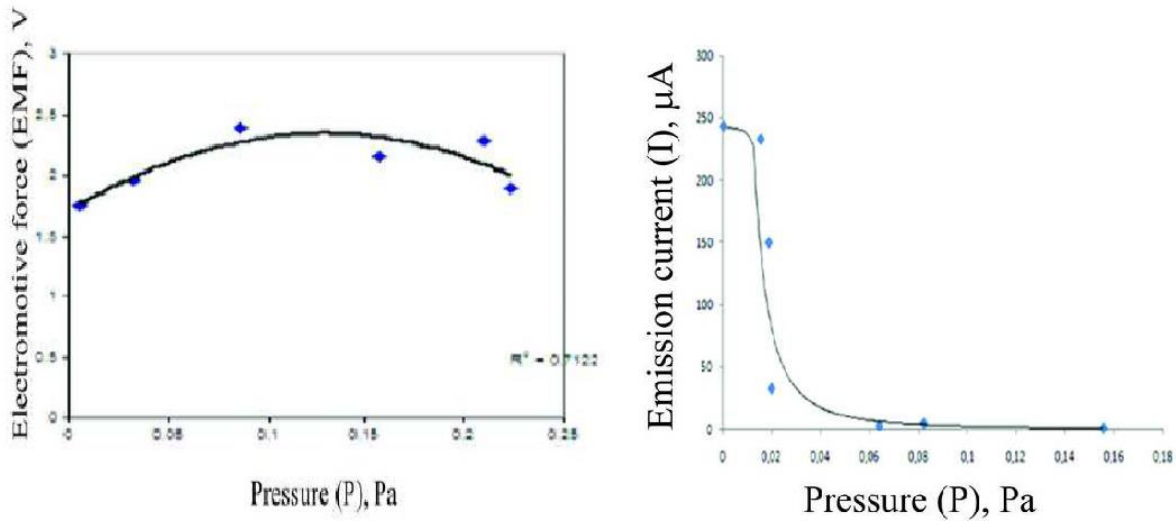

Fig. 2. Dependence of electromotive force versus a pressure (a) and emission current versus pressure of residual gas in a vacuum chamber (b).

In Fig. 2a, EMF measured under concentrated solar radiation is shown. It is seen from the linear dependence of EMF that there is the maximum of 2,3 V at the chamber pressure of 10-20 Pa.

It should be noted that even at atmospheric pressure a value of EMF is attained 0,5 V for Al-Li anode. However, emission current is too low that should be expected from operating at such pressure levels. So, it was not recorded in the mode of short circuit using conventional electrical measuring instruments (Fig. 2b).

Fig. 2b. shows an emission current versus pressure of residual gas in the vacuum chamber. Under concentration, a short-circuit current increases with decreasing chamber pressure. The highest experimental emission current of MWCNT with higher dislocation density is about 8 mA. This prototype system demonstrated 20 mW/cm$^2$ power conversion at standard conditions.

## 3. Basic Science

The most important factors in determining the solar energy conversion into electricity using MWCNT are enhanced absorption of solar radiation by MWCNT, excitation of its electronic subsystem and electronic transitions when electrons are excited from one energy level to a another. The field is enhanced near tips of carbon nano-tubes. The cations and electrons appear as a result of ionization of gas between electrodes under concentrated solar radiation.

In the electric field the positively charged particles move to the cathode, and electrons - to the anode. Effect of nano-tube tips on the cations which approaches to the surface of the cathode composed of nano-tubes increases in the growing electric field. The field reduces the barrier height (effect of Schottky) to the de Broglie wave-length ($\lambda \approx h/p$).

The mechanism of electron tunneling begins when cations get closer and electrons randomly jump from the surface of a cathode. It results in their neutralization, after that the forces of mirror reflection disappear, and the contact electric field stops to operate. Then weak Van der Waals forces are appeared. At T > 600 °C the desorption of neutral particles from a surface in the inter-electrode space occurs. As a result, emission current grows under the action of temperature and UV radiation. Voltage difference between electrodes is defined by the of works function of the cathode and anode. A process repeats itself again and again govern the anode current.

The metallic or silicon nano-wire of the cathode can function as a receiving antenna and energy transfer element when subjected to incident optical or infrared radiation. There is a displacement current, rather than flow of charge. Also, electrons of CNT more effectively move from the place of photon absorption to the surface (they travel between layers in nano-tubes as in waveguide).

## 4. Results and discussion

In the previous work, several mechanisms of electronic emission of carbon nano-tubes under the solar radiation were considered: As there is absorption of photons by molecules of residua atmosphere in a vacuum chamber under concentrated solar radiation, the molecule absorbing the photon energy is dissociated with formation of two particles. Each of the particles can be mainly, excited or in ionized state. The photo absorption mechanisms indicate that the photon's energy decrease is close to the energy of visible light.

The second mechanism of increase of concentration of charge carriers and emission current is photoelectron emission from the surface of carbon nano-tubes under solar radiation. In comparison with metals, CNT more effectively absorb solar radiation, and electrons of CNT can be more effectively excited because they are localized in direction of size quantization.

Although CNT can have outstanding electrical field emission properties (high emission currents at low electric field strengths), it is well known that the high emission capability of a single nano-tube does not necessarily translate directly into high emission current from a sample containing many nano-tubes because of the electrostatic screening effect.

Based on the experiments with non optimized components, a novel method of coupling solar cells to the CSP is proposed. To overcome space charge effects and to increase conversion efficiency, a conformal gate electrode with spacing <100 nm over 1 cm$^2$ area could be developed. The TFE unit prototype with the secondary mirror made of multilayer thin film filters and cathode based on carbon nano-tubes, hexaboride lantana, aluminum nitride with boron nitride, Ag-O-Cs and Ba-Sr-Ca oxide is designed and analyzed.

## 5. Future Work

The drawback of a thermionic energy (TE) converter is that it converts at only 20% efficiency and only at a very high temperature of about 1500$^0$ C. It is proposed the replacement of thermionic energy (TE) convertor with thermal field emission (TFE) convertor that would operate at about 40% efficiency and at much lower temperatures. Several factors that contributed to the relatively low TE efficiency could be corrected in the TFE design.

The proposed solar power system includes primary and secondary mirrors, optical reflector-anode, photo-enhanced gate and TFE elements directly coupled to capacitive storage. The movement of electrons between the emitter (cathode) and the collector (anode) is controlled by the gate (grid) serving as an electron generator itself. Also, gas having an ionizing potential which is intermediate between the gate and emitter work function potentials can be used. The PV and TFE modules placed in a glass envelope or vacuum tube allow to utilize an existing dish or PV-thermal (PVT) micro-concentrator to achieve the high conversion efficiency [7].

The initial designs are based on the existing SE or PVT rooftop concentrator from Chromasun Inc. and lithiated nano-particle diamond energy converter (LDFC) from University of Bristol. A dichroic mirror thermally decouples the modules by directing visible into commercial PV cells and ultraviolet-infrared (UV-IR) light into the TFE, LDFC or thermal receiver. (see Fig. 3).

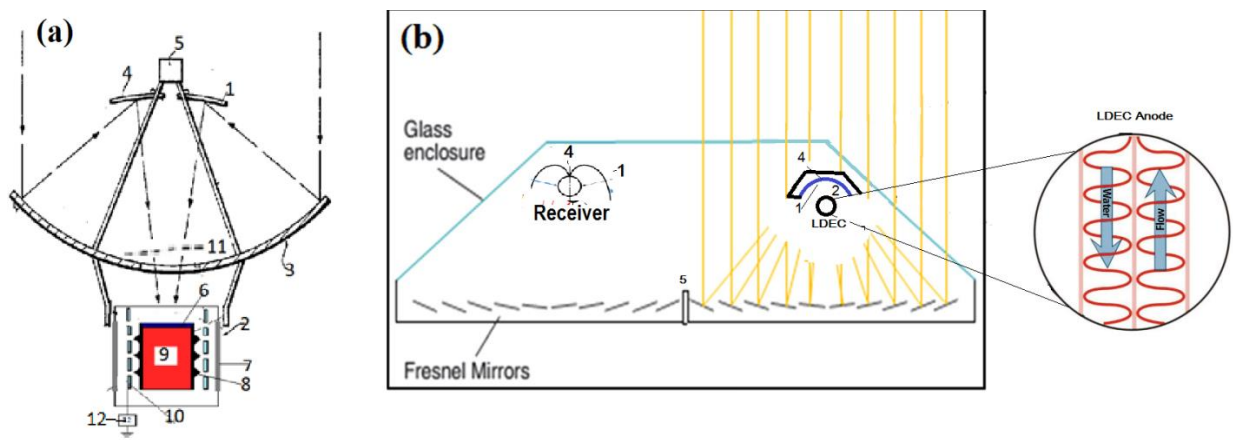

FIG. 3. a) . a) Hybrid dish and b) PVT units: 1 - dichroic mirror, 2 - TFE or LDFC module, 3 - primary mirror, 4 - PV cells, 5 - RF antenna and signal receptor, 6 - absorber, 7 - anode, 8 - cathode, 9 - thermal storage, 10 - gate, 11 - wire mesh, 12 - capacitive storage.

Also, dispersive lenses or perovskite thin-film solar cells with a light-trapping structure (LTS) can be used. The LTS idea is based on multi-frequency chessboard-like leaky modes in arrays of specially shaped metal nano-elements, i.e. double-bowtie antennas, that act as the spectral filter. The thin-film cells generate electricity from visible and reflect IR light into the absorber [8].

In the communication mode, radio frequency (RF) energy is redirected by an absorber or wire mesh to the RF antenna and signal receptor. The cooling fluid is boiled when cooling the PV cells and LDFC or TFE anode, and superheated at the thermal receiver. Quantum efficiencies of the TFE cathodes will be further enhanced in designs with more stable, low work-function Ba-Sr-Ca oxide, Ag or Ag-O-Cs coatings. To precise change various electrical properties of the material and simplify the design, a metal-vacuum-insulator-metal tunnel junction has been developed by researchers at Ilia State University (ISU), Georgia. The ISU team proposed a new method for promoting the passage of elementary particles at or through a potential barrier, having a geometrical shape for causing de Broglie interference between elementary particles [9].

The advanced design utilizes the high energy density capacitive storage, TFE cells with graphene carbon nano-tube (g-CNT) or planar edge cathodes and high transmission gates. To verify the simulation results, laboratory test will be done using the solar simulator. The concept behind the field tests is to compare the usable amount of energy directed from the beam splitters onto the TFE and PV modules to their direct solar exposure. In the tunneling photo-effect enhanced TFE cell prototype, the cathode (8) produces electrons and the gates (10) extract them. The sample holders have the plug-in for heating and measurement (Fig. 4).

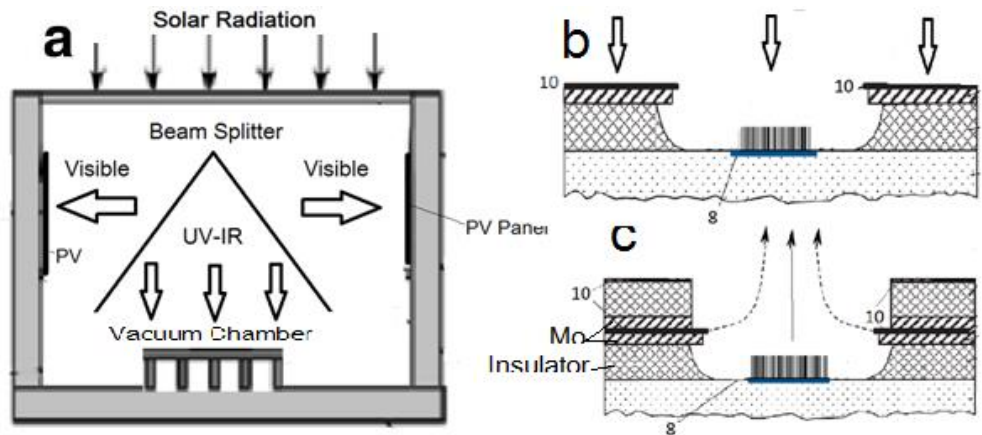

Fig. 4. Simplified configuration of (a) test unit, (b) single and (c) double-gate TFE cells.

Composite electrode materials will be developed by regulating adhesion between thin metal films to produce a large-area vacuum nanogap for maximum thermionic emission and thermo-tunneling. Then the sandwich will be separated and a vacuum nano-gap will be opened to produce two conformal electrodes. Ba-Sr-Ca oxide or Ag-O-Cs nano-grating surface cathode materials will be used to decrease the effective potential barrier between metal and vacuum.

Due to their low work function of 0.7 to 0.9 eV, Ag-O-Cs electrodes are highly sensitive to light, including near-infrared rays. The electrode surfaces will be analyzed by the Michelson interferometer. Both the adhesion strength and electroplating regime will be adjusted to obtain conformal electrodes with a large-area vacuum nano-gap for tunnel junction applications. Development of densely packed periodic structures and nano-grating has demonstrated changes in the electronic and electron emission properties of materials when the grating pitch becomes comparable with the de Broglie wavelength of the electron.

Such changes are due to the special boundary conditions imposed by nano-grating on the electronic wave function. The gate electrode will be fabricated on top of the nano-emitter arrays. The ISU technology allows us to make small structures on the surface of materials with the surface roughness less than the electron's wavelength.

For several years the Cherenkov Telescope Array (CTA) concept had been tested at the Sandia's National Solar Thermal Test Facilities and the previous experimental results of a two-mirror antenna design and modeling will be used to confirm the basic physics and equipment requirements for testing of the proposed hybrid beam-down concept. A non-imaging secondary (NIS) mirror comprised of concave or convex shapes will be analyzed. In addition to a NIS Fresnel lens, the NIS mirror originated to overcome the dark spot produced on a receiver by the secondary shadow can offer additional usage by preserving system efficiency.

A time synchronization is essential to enhance reliability of two-way electrical networks by better correlation in time of distributed energy resources and multiple loads. An Ethernet-based White Rabbit system synchronizes distributed data acquisition networks with high accuracy over fiber lengths of up to 10 km. During day time, an array of Cherenkov telescopes nightly operating in Arizona can serve as a precursor of the solar power micro-grid.

The system could be also used to analyze the systematic effects in the information or energy transfer between the power system and grid. Particularly, the CTA setup that utilizes time data from GPS satellites can be used to estimate with high speed and accuracy the frequency and power factors of the transmission system. In addition to GPS, a setup with the linked antennas and radio stars that serve as a natural frequency standard can be used to analyze the frequency and power factors of the long-distance transmission system.

**Conclusion**

Recently, several designs have been proposed to achieve spectral separation, an extensive review of which can be found in [10]. While these concepts have proven to be effective for spectral splitting, they are not suitable for low-cost industrial scale manufacturing. In this article we describe the preliminary study of a thermo emission conversion device use in the CSP with beam splitting to fulfill the requirements of mass production. At first, the preliminary examination of thermal energy conversion into electricity under solar radiation heating was carried out.

Electrodes for solar-to-electrical energy converters based on homogeneous mixture of MWCNT of high stiffness, strength and resilience and flexible expanded graphite were also produced and investigated. Mechanisms of electronic emission of nano-structured electrodes and process of gas ionization under the influence of solar radiation at different intensities, temperatures and vacuum conditions were measured.

The design has great commercial potential for applications in remote areas with a great solar resource, particularly in space, military and astrophysics installations. For large scale applications, a high-efficiency CSP with the low average electricity price is possible using the TFE module. It can be modeled by a high-current electron gun that uses a thoriated tungsten filament. Thorium radioactivity is too small to rule it out in devices that are used to explore the thermionic fission electric cell-based transmutation reactors, neutron and X-ray sources [11].

Now, hybrid battery and super-capacitor (SC) energy storages buffer large and rapid fluctuations in energy supply and demand. approach to ensure system reliability and autonomy. Since thermal energy storage is cheap, a hybrid SC and thermal storage-integrated TFE module can be much less expensive approach [12].

**Acknowledgements**

The authors wish to acknowledge support of CRDF (UKE2-7034-KV-11).